\documentstyle[prl,preprint,aps,epsf]{revtex}

\begin{document}
%\draft
\title {Statistical spectroscopic calculation of expectation 
values and spin-cutoff factors}
\author{Calvin W.~Johnson$^1$, Jameel-Un Nabi$^1$ and W.~Erich Ormand$^2$ }
\address{
$^1$Department of Physics and Astronomy\\
Louisiana State University,
Baton Rouge, LA 70803-4001\\
$^2$Lawrence Livermore National Laboratory\\
P.O. Box 808, Mail Stop L-414 \\
Livermore, CA 94551
}

\maketitle
\begin{abstract}
Recently we proposed a formalism, based in nuclear 
statistical spectroscopy, for efficient computation of 
nuclear level density, or densities of states,  
through a sum of partitioned binomial functions  (SUPARB).  
In this Letter we extend the formalism 
to the calculation of locally averaged expectation values, 
with specific application to spin-cutoff factors and the angular 
momentum dependence of the nuclear density of states. 
\end{abstract}
%\copyright{2001 American Physical Society}


In a recent Letter \cite{Joh01} 
we proposed a computationally efficient method to compute 
nuclear level densities, based in the ideas of 
nuclear statistical spectroscopy \cite{Mon75,Fre83,Won86}.   While we 
refer interested readers to our original Letter for details, we briefly 
summarize 
here. Consider the density of states  of a  
Hamiltonian $\hat{H}$, 
\begin{equation}
\rho(E) = {\rm tr \, } \delta (E- \hat{H}).
\end{equation}
It is useful to partition the finite model space into subspaces, 
labeled by Greek letters $\alpha, \beta$, etc., each with an associated 
projection operator $P_\alpha$.  Then one can define {\it partial} 
or configuration densities:
\begin{equation}
\rho_\alpha(E) = {\rm tr \, } P_\alpha \delta (E- \hat{H}).
\end{equation}
The total density is the sum of the partial densities. 
(NB:We {\it always} include  
$2J+1$ degeneracies and so are formally considering state densities.) 

 If the subspaces are single-particle configurations, e.g., $(0d_{5/2})^4$, 
$(0d_{5/2})^2(1s_{1/2})^2$, etc, then the partial, or configuration,
moments up to fourth order for any system with $Z$ protons and $N$ 
neutrons in the valence space can be computed directly from the 
one+two-body matrix elements of $\hat{H}$ \cite{Fre71,Ayi74}.  
For any partition (configuration) $\alpha$ let 
$d_\alpha = {\rm tr \,} P_\alpha$ be the dimension of the 
subspace $\alpha$, and define the configuration average over the 
subspace to be  
$ \langle \ldots \rangle_\alpha \equiv 
{ d_\alpha}^{-1} {\rm tr \,} P_\alpha ( \ldots ) 
$.
Then 
 $\bar{E}(\alpha) = \langle \hat{H}\rangle_\alpha$
is the (configuration) centroid, 
$\gamma(\alpha) = \langle (\hat{H} - \bar{E}_\alpha)^2 \rangle_\alpha^{1/2}$
  the configuration width,  
$m_3 (\alpha) =  
\langle (\hat{H} - \bar{E}_\alpha)^3 \rangle_\alpha /\gamma^3(\alpha)$  
the scaled
 (dimensionless) third configuration moment, and 
$m_4 (\alpha)=   \langle (\hat{H} - \bar{E}_\alpha)^4 \rangle_\alpha
/\gamma^4 (\alpha)$ the  scaled fourth configuration moment. 

With these moments in hand, we model the partial  densities as 
binomial distributions, 
following a recent suggestion of Zuker \cite{Zuk01}.  
Starting with the binomial expansion of 
$(1+\lambda)^N $ and representing ${N \choose k}$ with gamma functions, one 
can derives a continuous distribution,
\begin{equation}
\rho(E_x) = \lambda^{E_x / \epsilon} 
{ \Gamma(E_{max}/\epsilon+1) \over \Gamma(E_x / \epsilon +1) 
\Gamma( (E_{max} - E_x) /\epsilon +1 )}
\label{density}
\end{equation}
where $E_x$ is the excitation energy and 
$E_{max} = \gamma (1 + \lambda) \sqrt{N/\lambda}$ is the maximum 
excitation energy in the binomial distribution. 
The binomial is appealing 
because one can easily compute the scaled third and fourth central moments: 
$ m_{3}={1- \lambda \over \sqrt{N \lambda}}$ and 
$ m_4 =
3- {4-\lambda \over N}+ {1 \over N \lambda},$
allowing one to control the shape of the binomial through $N$ and $\lambda$.

To compute the density of states, we take the following steps: 
(1) We compute the configuration moments up to 3rd or 4th order. 
(2) We model the partial density  for each configuration 
as a binomial. The binomial 
parameters $N$ and $\lambda$, as well as the overall energy scale 
and centroid, are fitted to the configuration moments. (3) The 
partial densities are summed to yield the total density of states. 
Because of these ingredients, we will refer to our approach as 
SUPARB (SUm of PARtitioned Binomials) state densities. We have shown 
that our method models well the density of states, when compared against 
exact calculations, and that 
one needs third configuration moments and occasionally, but not 
always, fourth configuration moments to achieve accurate results \cite{Joh01}. 
Incidentally, none of this
is special to the atomic nucleus. One could easily apply it to atomic 
electrons as well.

We now turn to the computation of expectation values of some general operators,
 $\hat{\cal O}$, and define the energy-dependent 
 locally averaged expectation value (LEV) as
\begin{equation}
\langle \hat{\cal O} (E) \rangle
= { {\rm tr \,}\hat{\cal O} \delta(E-\hat{H})   \over
{\rm tr \,} \delta(E-\hat{H})   }.
\end{equation}
We also have configuration LEVs:
\begin{equation}
\langle \hat{\cal O} (E) \rangle_\alpha
= { {\rm tr \,}P_\alpha \hat{\cal O} \delta(E-\hat{H})   \over
{\rm tr \,} P_\alpha\delta(E-\hat{H})   }.
\label{LEVconfig}
\end{equation}

Our strategy for computing the total $\langle \hat{\cal O} (E) \rangle$
is the same as for the total density of states. 
In addition to the configuration moments, we compute the 
weighted averages
$\langle \hat{\cal O} \rangle_\alpha$,
 $\langle \hat{\cal O}(\hat{H}-\bar{E}_\alpha) \rangle_\alpha$,
and  $\langle \hat{\cal O}(\hat{H}-\bar{E_\alpha})^2 \rangle_\alpha$.
These can be computed directly from one+two-body matrix elements 
similar to those for the moments \cite{Fre71,Ayi74}.

The only tricky point is if 
 $\hat{\cal O}$ and $\hat{H}$ do not commute; 
where does one insert the projection operator $P_\alpha$ in 
eqn.~(\ref{LEVconfig})? 
Because $\sum_\alpha P_\alpha = 1$ by the completeness of the
projection operators, we conclude consistency is the only requirement,
and take 
${\rm tr \,}  P_\alpha \hat{\cal O} \hat{H}^k$, $k = 0, 1, 2$.

Assume within any subspace a quadratic energy dependence, that is, 
\begin{equation}
\langle \hat{\cal O} (E) \rangle    =
{\cal O}_0 +    {\cal O}_1  (E-\bar{E})/\gamma + {\cal O}_2   
(E-\bar{E})^2/\gamma ^2
\end{equation}
(dropping for the moment the subspace label $\alpha$). 
Then
\begin{eqnarray}
\langle\hat{\cal O} \rangle =    {\cal O}_0+ {\cal O}_2, \\
\langle\hat{\cal O}(\hat{H}-\bar{E}) \rangle =  \gamma (   {\cal O}_1+ {\cal O}_2 m_3),  \\
\langle\hat{\cal O}(\hat{H}-\bar{E})^2 \rangle =  \gamma^2 (   {\cal O}_0+
{\cal O}_1 m_3+ {\cal O}_2 m_4  ).
\end{eqnarray}
Solving for coefficients,
\begin{eqnarray}
{\cal O}_0 = {
 \langle\hat{\cal O} \rangle (m_4-m_3^2) +  m_3\langle\hat{\cal O}(\hat{H}-\bar{E}) \rangle /\gamma
 -  \langle\hat{\cal O}(\hat{H}-\bar{E})^2 \rangle /\gamma^2
\over
m_4 - m_3^2 -1 }, \\
{\cal O}_1 = {
   (m_4 -1) \langle\hat{\cal O}(\hat{H}-\bar{E}) \rangle /\gamma
 + m_3 \left (   \langle\hat{\cal O}(\hat{H}-\bar{E})^2 \rangle /\gamma^2  -\langle\hat{\cal O} \rangle
 \right )
\over
m_4 - m_3^2 -1 }, \\
{\cal O}_2 = {
\langle\hat{\cal O}(\hat{H}-\bar{E})^2 \rangle /\gamma^2
  -  m_3\langle\hat{\cal O}(\hat{H}-\bar{E}) \rangle /\gamma - \langle\hat{\cal O} \rangle
\over
m_4 - m_3^2 -1 }
\end{eqnarray}
If we limit ourselves to only a linear dependence (assume ${\cal O}_2 = 0$), 
then
\begin{eqnarray}
{\cal O}_0 =         \langle\hat{\cal O} \rangle, \\
{\cal O}_1 = \langle\hat{\cal O}(\hat{H}-\bar{E}) \rangle /\gamma.
\end{eqnarray}
Note: in terms of the binomial parameters $N, \lambda$,   one finds that
$m_4 -m_3^2 -1 = 2(1-N^{-1})$.

The generalization to the  full SUPARB case, summing over partitions, 
is easy:
\begin{equation}
\langle \hat{\cal O} (E) \rangle
= {
\sum_\alpha     \langle \hat{\cal O} (E) \rangle_\alpha \rho_\alpha(E)
\over
\rho(E)         }
\end{equation}

In Figures 1-3 we illustrate our method, comparing to ``exact'' 
shell model calculations in full $0\hbar\omega$ spaces. 
 For $sd$-shell nuclides we used 
the Wildenthal USD interaction \cite{Wil84}, and compared against direct 
diagonalization. Fig.~1 shows the LEV of $\vec{Q}\cdot \vec{Q}$ 
and $S^2$ for $^{20}$Ne, while Fig.~2 shows the LEV of $J^2$ 
(which will be important for the spin-cutoff factor below) 
for $^{22}$Na, $^{23}$Mg, and $^{32}$S. 

Fig.~3 compares the LEV of $J^2$ for $^{48}$Cr and $^{54}$Fe 
in a full $0\hbar\omega$ $pf$-shell calculation; here the 
 ``exact'' calculation 
was through Monte Carlo sampling of path integrals\cite{Joh92,Orm97}. 
To avoid the well-known sign problem \cite{Dea95}  we fitted a schematic 
multipole-multipole interaction  to the $T=1$ matrix elements 
of the FPD6 interaction of Richter {\it et al.} \cite{richter}.  
Clearly the full SUPARB calculation, with a quadratic dependence in 
each partition, works very well, and the quadratic is a significant 
improvement over the linear approximation.  One could go to a cubic 
dependence, but computation of the necessary moments, while possible, 
would be very time consuming and there is not much room for improvement.

Now we apply our formalism to spin-cutoff factors for state densities.  
Define the $J$-dependent density as
  \begin{equation}
\rho(E,J) = {\rm tr \,}\left (  \delta(E-\hat{H})   \delta(J(J+1)-\hat{J}^2)
\right ) .
\end{equation}
Traditionally this is factorized to 
\begin{equation}
              \rho(E,J) = \rho(E) (2J+1)\Omega(J,E),
\end{equation}
and  one assumes a weighted Gaussian form for $\Omega$:
\begin{equation}
\Omega(J,E) = (2J+1) { 1 \over 4 \sigma(E) \sqrt{2 \pi} }
\exp \left( - { (J+ {1 \over 2} )^2 \over 2 \sigma(E)^2 } \right )
\label{spincutoff}
\end{equation}
We have normalized $\int dJ \Omega(J,E) = 1$.
Here $\sigma(E)$ is the ``spin-cutoff'' factor and is  dependent
on the energy. The spin-cutoff factor has been considered before 
in the context of nuclear statistical spectroscopy \cite{Gri83}, but 
the moments were computed by a random sampling of representative 
vectors and the level densities were approximated by Hermite polynomials 
(which do not guarantee nonnegative densities). 

Given the form (\ref{spincutoff}) one finds that 
\begin{equation}
\langle \hat{J}^2 \rangle = 3 \sigma^2 - {1 \over 4}
\end{equation}
(The factor of 3 is because we must include the $2J+1$ degeneracy in 
our traces.)
Therefore, the problem of describing $\rho(E,J)$ with SUPARB reduces to
computing $\langle \hat{J}^2 \rangle $ as a function of energy, 
which were given in figures 2 and 3. 

In figure 4 we plot the exact and SUPARB $J$-projected level densities 
(without the $2J+1$ degeneracy) of $^{32}$S, for $J= 0,1,2$ and 8. 
The results are very good.  Incidentally, one could compute 
the spin-cutoff factors and thus the $J$-projected level densities 
using Monte Carlo evaluation of path integrals just as easily, 
although this has not yet been done. 

In summary, we have extended our previous SUPARB technique for 
state densities to expectation values of operators, including application 
to spin-cutoff factors. Our results look very good. Other possible 
application include estimating the contamination of spurious states in 
cross-shell calculations, and computation of total strengths and 
energy-weighted sum rules for transitions.

This work was performed under the auspices of the 
Louisiana Board of Regents, 
contract number LEQSF(1999-02)-RD-A-06; and under the auspices of the U.S. Department of Energy 
through the University of California, Lawrence Livermore National Laboratory, 
under contract No. W-7405-Eng-48.

\begin{figure}[h!]
\epsfxsize=10.75cm	%this is size for preprint
\epsffile{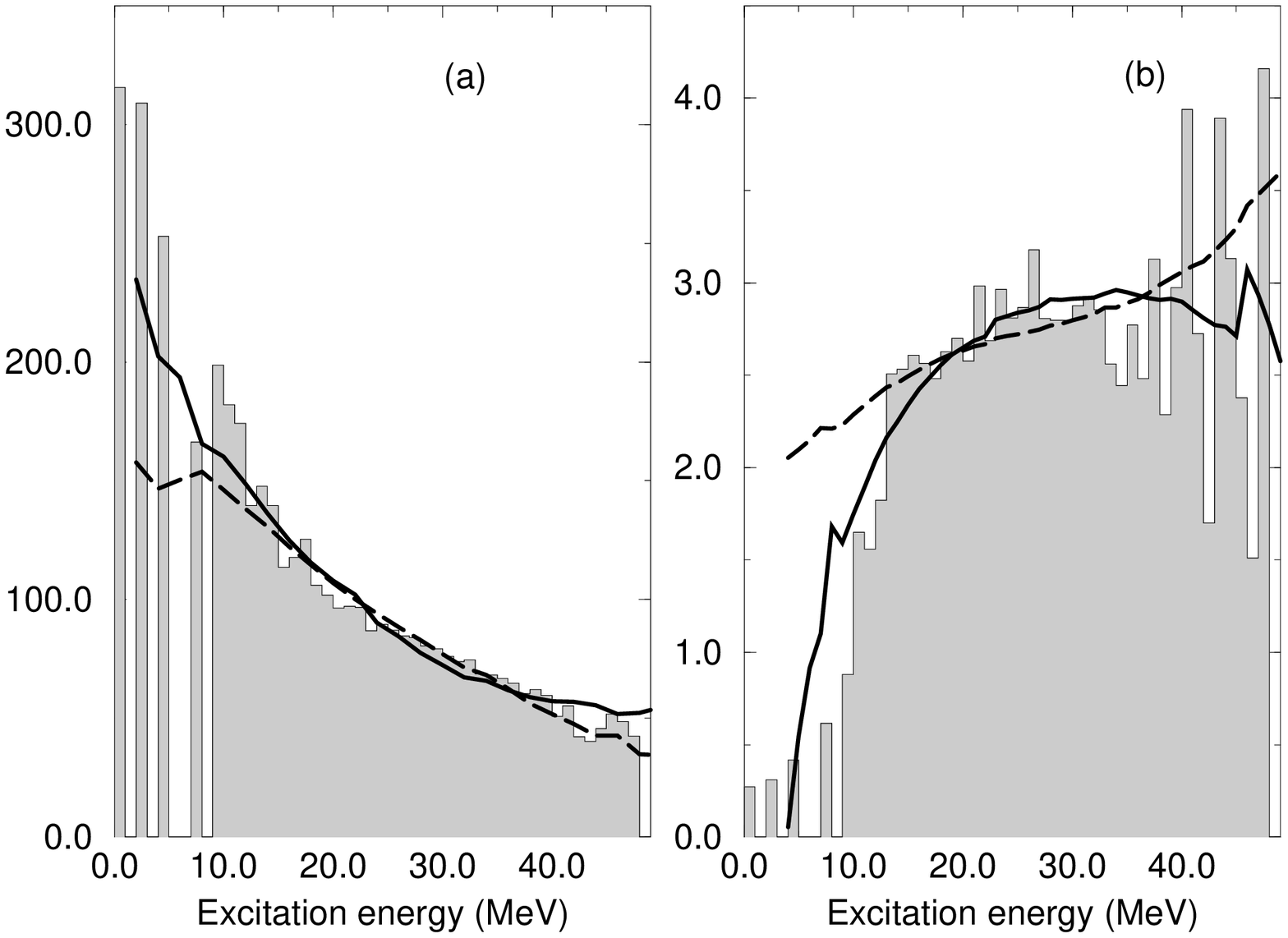}
\caption{ Comparison of exact shell model locally-averaged 
expectation value of (a) $\vec{Q} \cdot \vec{Q}$ and (b) $S^2$ 
in $^{20}$Ne. The histogram is  
from direct diagonalization, and we compare against 
both linear (dashed line) and quadratic (solid line) SUPARB estimates. 
}
\end{figure}

\begin{figure}[h!]
\epsfxsize=10.75cm	%this is size for preprint
\epsffile{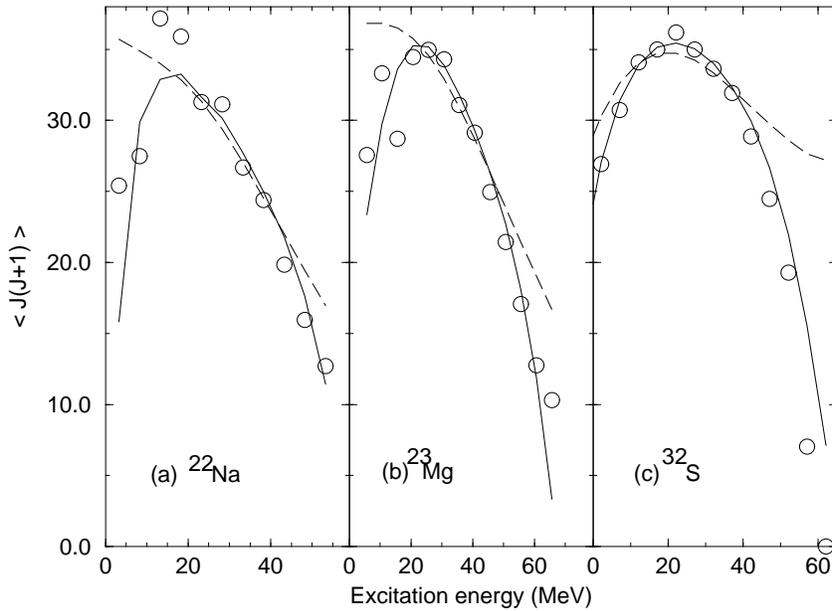}
\caption{ Comparison of exact shell model locally-averaged 
expectation value of $J^2$ in (a) $^{22}$Na, (b) $^{23}$Mg, 
and (c) $^{32}$S.  The circles are 
from direct diagonalization, and we compare against 
both linear (dashed line) and quadratic (solid line) SUPARB estimates. 
}
\end{figure}

\begin{figure}[h!]
\epsfxsize=10.75cm	%this is size for preprint
\epsffile{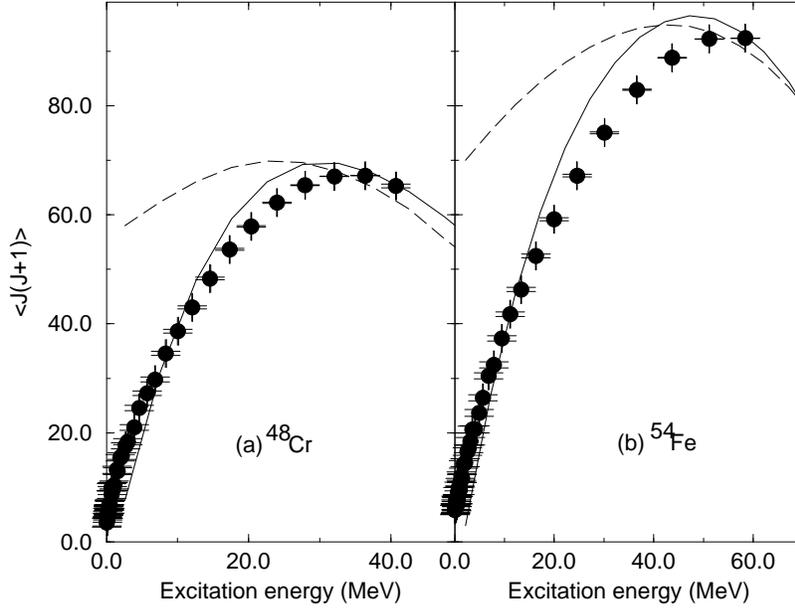}
\caption{ Comparison of exact shell model locally-averaged 
expectation value of $J^2$ in (a) $^{48}$Cr, 
and (b) $^{54}$Fe.  The histogram is  
from Monte Carlo sampling of a path integral, and we compared against 
both linear (dashed line) and quadratic SUPARB  (solid line) estimates. 
}
\end{figure}

\begin{figure}[h!]
\epsfxsize=10.75cm	%this is size for preprint
\epsffile{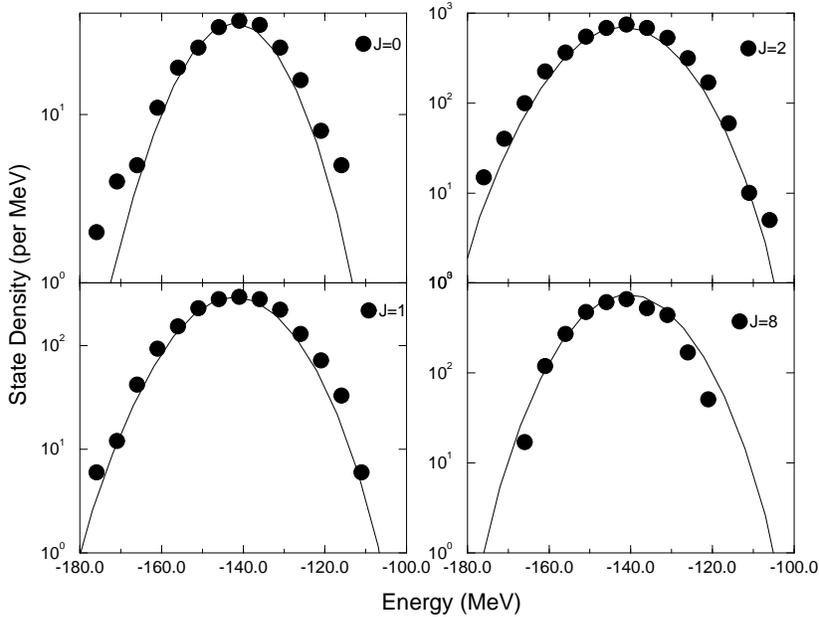}
\caption{ Comparison of $J$-projected state densities 
in $^{32}$S for J = 0, 1, 2, and 8. The circles are `exact'   
from direct diagonalization, and the solid lines 
are SUPARB estimates with spin-cutoff factor. 
}
\end{figure}

\end{document}